\documentclass[twocolumn,tighten]{aastex63}
\usepackage{amssymb, amsmath,framed,chemfig}

\usepackage{bm}
\expandafter\ifx\csname package@font\endcsname\relax\else
 \expandafter\expandafter
 \expandafter\usepackage
 \expandafter\expandafter
 \expandafter{\csname package@font\endcsname}
\fi
\def\bq{\begin{equation}}
\def\eq{\end{equation}}
\def\bqy{\begin{eqnarray}}
\def\eqy{\end{eqnarray}}
\def\p{\partial}

\def\p{\partial}

\def\R{\mathbb{R}}

 \def\q#1#2{q^{\hspace{1pt}#1}_{\,,\hspace{1pt}#2}}
 \def\a#1#2{a^{\hspace{1pt}#1}_{\,,\hspace{1pt}#2}}
 \def\d#1#2{\delta^{\hspace{1pt}#1}_{\,,\hspace{1pt}#2}}

\begin{document}
\title{\large{The Extended Habitable Epoch of the Universe for Liquids Other than Water}}

\correspondingauthor{Manasvi Lingam}
\email{mlingam@fit.edu}

\author{Manasvi Lingam}
\affiliation{Department of Aerospace, Physics and Space Sciences, Florida Institute of Technology, Melbourne FL 32901, USA}
\affiliation{Institute for Theory and Computation, Harvard University, Cambridge MA 02138, USA}

\author{Abraham Loeb}
\affiliation{Institute for Theory and Computation, Harvard University, Cambridge MA 02138, USA}

\begin{abstract}
At high redshifts, the temperature of the cosmic microwave background (CMB) was higher than its value today. We explore the possibility that life may have arisen early because the higher CMB temperature would have supplied the requisite energy for the existence of different solvents on the surfaces of objects. At redshifts of $z \lesssim 70$, after the first stars are predicted to have formed, a number of molecules (but not water) might have existed in liquid form over intervals of $\sim 10$ Myr to $\sim 100$ Myr. We delineate the challenges and prospects for life in the high-redshift Universe, and assess the various candidates for alternative biochemistries in this context -- of the options considered herein, we conclude that ethane is probably the most promising contender. \\
\end{abstract}

% Linenumbers required for AAS journals; use "linenumbers" in the documentclass command, at the beginning of the manuscript.

\section{Introduction}\label{SecIntro}
The habitable zone (HZ) is defined as the region around a star where liquid water can exist on the surface of a planet \citep{KWR93}. This concept has been similarly extended to encompass solvents other than water \citep[e.g.,][]{BFM}. It is, however, essential to recognize that regions other than the HZ may comprise objects with temperatures amenable to the presence of solvents. Notable examples include icy moons with subsurface oceans, worlds with deep biospheres, and free-floating objects with thick atmospheres \citep{SMI18,LL21}. 

There is however, another possibility that merits explication. In the early Universe, the cosmic microwave background (CMB) had a higher blackbody temperature \citep{LF13}. At a redshift of $z \sim 100$, which corresponds to a Universe that was merely $\sim 16.6$ Myr old, the CMB temperature was around $275$ K. Hence, if any planets existed at this time, they would have possessed surface temperatures conducive to liquid water due to the energy supplied by the CMB. In this scenario described by \citet{Lo14}, the distance from a star would be rendered irrelevant and virtually all planets would have clement temperatures.

However, the potential drawback is that the first stars are anticipated to have formed at redshifts of $z \lesssim 70$ as per numerical models \citep{Bar16,Lo20},\footnote{There is tentative observational support for stars at $z \sim 20$ \citep{Bou18}, but the data is subject to ambiguities. The launch of the James Webb Space Telescope (JWST) is predicted to substantially advance our knowledge of the early Universe.} although rare star formation at redshifts of $z \sim 100$ is not impossible in theory \citep{Lo14}. By taking our cue from this consideration and by drawing upon the progress made in our understanding of ``exotic'' biochemistries, we explore the epochs at which the early Universe may have been amenable to the existence of non-water solvents on objects by virtue of the CMB.

The outline of this paper is as follows. In Sec. \ref{SecHW}, we outline the major alternatives to water as a solvent and calculate the corresponding ages at which they could have existed in liquid form. In Sec. \ref{SecLifeU}, we delve adumbrate the feasibility of life in the high-redshift Universe. Finally, we summarize our findings and outline the most promising candidates in Sec. \ref{SecConc}.

\section{Solvents and their respective habitability epochs}\label{SecHW}
In light of the preceding discussion, we will focus on chemical compounds that are liquids at temperatures conspicuously lower than the standard range of $273$-$373$ K for liquid water at a pressure of $1$ bar. The reason that we select solvents with melting points of $T < 273$ K is because the CMB temperature $T_\mathrm{CMB}$ obeys
\begin{equation}\label{TCMB}
    T_\mathrm{CMB} \approx 2.73\,\mathrm{K}\,(1 + z).
\end{equation}
Hence, if the solvent has a melting point of $T > 273$ K, it would imply that the corresponding redshift to yield this ambient CMB temperature is characterized by $z > 100$. This poses difficulties because the first stars are expected to have formed only at redshifts of $z \lesssim 70$ \citep{LF13,Lo14,Bar16}.

There are a number of alternative solvents that have been discussed in the literature with regards to ``exotic'' life. On the one hand, there are compelling physicochemical grounds as to why water constitutes an excellent solvent for the origin and sustenance of life \citep{PP12,WB18}; the inherent benefits include, \emph{inter alia}, its high heat capacity, its large thermal range over which it remains liquid, its capability to dissolve a diverse array of compounds, and its anomalous thermal expansion. On the other hand, there are no definitive reasons that appear to rule out the possibility of solvents other than water. 

Hence, we will tackle a few representative examples of chemical species that are liquid at temperatures $T < 273$ K. Our list is \emph{not} meant to be comprehensive; note that in-depth analyses of this topic have been furnished in \citet{Fir63,Ba04,BRC04,SMI18}. In fact, the abundances of oceans composed of certain non-water solvents (e.g., ethane, methane, and nitrogen) might be several times higher than their water counterparts in the current Universe, as elucidated in \citet{BFM}.

There are a couple of points that merit explication before proceeding further. First, we conservatively adopt the melting ($T_m$) and boiling ($T_b$) temperatures at the standard pressure (SP) of $1$ bar. At higher pressures, $T_b$ could increase significantly until the critical temperature is reached, but this would correspond to a higher redshift as seen from (\ref{TCMB}), which decreases the likelihood for the existence of stars and planets. Second, we will not consider mixtures comprising multiple solvents because the corresponding values of $T_m$ and $T_b$ ought to overlap with the chemical species that make up the mixtures; in a similar vein, we shall not tackle solvents with substantial concentrations of solutes (e.g., brines). 

The most commonly studied alternative to water in the context of carbon-based biochemistry is ammonia (NH$_3$). It has been theorized since at least the 1950s and 1960s that ammonia may constitute the basis for an alternative biochemistry \citep{Hal54,Fir63}. The replacement of the C$=$O bond in the carbonyl group with the C$=$N bond could give rise to a plausible metabolism \citep{BRC04}. The synthesis of peptide bonds from amino acids, which form the backbone of proteins, is rendered plausible with the release of NH$_3$ as the product \citep{Fir63}. At SP, the melting and boiling points of ammonia are respectively given by $T_m \approx 195.5$ K and $T_b \approx 239.8$ K.

Hydrogen sulfide (H$_2$S) has been posited as an alternative to water in the category of polar inorganic solvents. Although it has a lower dipole moment and exists in liquid form over a narrower temperature range than water, it is known to dissolve a sizable number of organic compounds and the SH$^-$ anion formed after dissociation is capable of playing a similar role to that of the hydroxyl ion in organic molecules \citep{SMI18}. Lastly, the abundance of H$_2$S seas is conceivably close to that of H$_2$O seas in the present day \citep[Figure 6]{BFM}. Hydrogen sulfide is characterized by $T_m \approx 187.7$ K and $T_b \approx 213$ K at SP. 

Next, we turn our attention to hydrocarbons. Despite their non-polar nature, there is ample empirical evidence suggesting that the organic reactivity of hydrocarbon solvents is not much lesser than water \citep{BRC04}. Although we could tackle any of the alkanes, we will focus on ethane (C$_2$H$_6$) for the following reasons: (i) it is liquid over a broad temperature range, (ii) it has a low boiling point, (iii) it has been detected regularly in the interstellar medium \citep{Co20}, (iv) it represents one of the major components of the lakes and seas of Titan \citep{McK16}, and (v) ethane seas may be more prevalent than water seas by an order of magnitude in the current Universe \citep[Figure 6]{BFM}. 

Polarity-inverted membranes for putative cells have been identified for hydrocarbons \citep{SLC15}, although some questions about their efficacy have subsequently arisen \citep{SR20}. It is plausible that polyethers could serve as genetic polymers in hydrocarbon solvents at least up to temperatures of $170$ K \citep{MOI15}, and other candidates have also been explored from a theoretical standpoint \citep{McK16}. In the case of ethane, we specify $T_m \approx 90.4$ K and $T_b \approx 184.2$ K at SP. Although propane (C$_3$H$_8$) is much less abundant than ethane in Solar system objects like Tian \citep{VBN10} and laboratory experiments simulating the physicochemical conditions of the interstellar medium \citep{Kai02}, we will consider this molecule for the sake of comparison; we adopt $T_m \approx 83.5$ K and $T_b \approx 231$ K at SP. Furthermore, propane has been deployed in a few experiments probing alternative biochemistries \citep{SLC15}.

Methanol (CH$_3$OH), a polar organic compound, was highlighted as a potential solvent by \citet{SMI18}. It has a high dipole moment and dielectric constants, remains in liquid form over a considerable thermal range, and is possibly better than water as a temperature moderator due to its high heat of vaporization. Moreover, it is relatively common in the Universe, as evinced by observations of this molecule in the interstellar medium \citep{Co20}. At SP, methanol has the properties of $T_m \approx 175.6$ K and $T_b \approx 337.9$ K. 

Molecular nitrogen (N$_2$) is an example of a non-polar inorganic solvent that may be more plentiful than water in its liquid form on planets and moons today \citep[Figure 6]{BFM}. Although liquid N$_2$ is not conducive to the emergence of life-as-we-know-it, \citet{Ba04} proposed that silanols (the silicon analogs of alcohols) could potentially dissolve in this solvent and thereby engender the precursors of exotic life. At SP, molecular nitrogen is liquid over the narrow thermal range of $T_m \approx 63.2$ K and $T_b \approx 77.4$ K. 

Supercritical fluids are worthy of consideration as heterodox solvents. This is particularly true of supercritical carbon dioxide since it has desirable (bio)chemical properties (e.g., stabilizing enzymes) and is documented to host life in habitats at the ocean floor on Earth \citep{BSM14}. However, as we are interested in solvents at temperatures much lower than $273$ K, we will consider the extreme case of supercritical dihydrogen (H$_2$); this was discussed as a prospective solvent in the atmospheres of the Jovian planets by \citet{BRC04}. The lower bound for supercritical H$_2$ is set by the critical point and translates to $T_m \approx 33.3$ K at a pressure of $13$ bar \citep[their Table 1]{BRC04}.\footnote{\url{https://www.engineeringtoolbox.com/hydrogen-d_1419.html}} At a pressure of $\mathcal{O}(10)$ bar, supercritical H$_2$ is feasible at temperatures up to $> 10^3$ K.\footnote{\url{https://commons.wikimedia.org/wiki/File:Phase_diagram_of_hydrogen.png}} Given that Super-Earths appear to have thick H$_2$ atmospheres for multiple reasons \citep{SHP20}, it is plausible that supercritical H$_2$ may exist in some regions of these planets. 

\begin{table*}
\begin{minipage}{165mm}
\caption{Habitability epochs of the Universe for various solvents}
\label{TableHabEp}
\vspace{0.1 in}
\begin{tabular}{|c|c|c|c|c|c|c|c|}
\hline
{\bf Solvent} & $T_m$ (in K) & $T_b$ (in K) & $z_m$ & $z_b$ & $\tau_m$ (in Myr) & $\tau_b$ (in Myr) & $\Delta t$ (in Myr) \tabularnewline
\hline
\hline
Ammonia & 195.5 & 239.8 & 70.6 & 86.8 & 28.1 & 20.6 & 7.5\tabularnewline
\hline
Hydrogen sulfide & 187.7 & 213 & 67.8 & 77.0 & 29.9 & 24.7 & 5.2\tabularnewline
\hline
Ethane & 90.4 & 184.2 & 32.1 & 66.5 & 90.7 & 30.8 & 59.9\tabularnewline
\hline
Propane & 83.5 & 231 & 29.6 & 83.6 & 103 & 21.8 & 81.2\tabularnewline
\hline
Methanol & 175.6 & 337.9 & 63.3 & 122.8 & 33.1 & 12.1 & 21.0\tabularnewline
\hline
Molecular nitrogen & 63.2 & 77.4 & 22.2 & 27.4 & 156 & 115 & 41\tabularnewline
\hline
Supercritical H$_2$ & 33.3 & $>$ 1000 & 11.2 & 365.3 & 409 & $<$ 2.2 & $\sim$ 400\tabularnewline
\hline
\end{tabular}

\medskip

\textbf{\textit{Additional notes:}} $T_m$ and $T_b$ are the melting and boiling point of the solvent at the standard pressure of $1$ bar, except for supercritical H$_2$ for which a pressure of $\mathcal{O}(10)$ bar is needed. $z_m$ and $z_b$ represent the corresponding redshifts determined via (\ref{TCMB}), whereas $\tau_m$ and $\tau_b$ are the ages of the Universe at these redshifts. $\Delta t = \tau_m - \tau_b$ signifies the duration over which the solvent could persist on the surface (i.e., habitability epoch).
\end{minipage}
\end{table*}

For the list of candidates assembled hitherto, it is straightforward to estimate the equivalent redshifts required for the CMB to have temperatures of $T_m$ and $T_b$, which are denoted by $z_m$ and $z_b$, respectively. Once the redshifts are known, it is straightforward to determine the ages of the Universe ($\tau_m$ and $\tau_b$) in the standard cosmological model by means of the calculator described in \citet{Wri06}.\footnote{\url{http://www.astro.ucla.edu/~wright/CosmoCalc.html}} The age of the Universe (denoted by $t$) as a function of the redshift $z$ \citep{LF13} during the matter-dominated era is:
\begin{equation}
   t \approx 30\,\mathrm{Myr}\,\left(\frac{1+z}{70}\right)^{-3/2}.
\end{equation}
The final results of our analysis are depicted in Table \ref{TableHabEp}. The most relevant parameter is $\Delta t$ because it embodies the duration of time over which a particular solvent may exist in liquid form on any object owing to the heating from the CMB. It functions as a proxy for the interval over which the chemical species can exist in liquid form, which we label the habitability epoch of the solvent.

\section{Prospects for life in the early Universe}\label{SecLifeU}
We will briefly evaluate the prospects for life in the high-redshift Universe, with the proviso that this analysis is permeated by many uncertainties and unknowns. \\

\noindent \textbf{Star and planet formation:} As briefly discussed in Sec. \ref{SecIntro}, theoretical models favor $z \lesssim 70$ for the origin of first stars, although it is possible that rare stars may have collapsed directly from halos at redshifts of $z \sim 100$ \citep{Lo14}. Once the first generation (Population III) stars formed, they had typical lifetimes of a few Myr \citep{BKL01,VB13}, after which they exploded as core-collapse or pair-instability supernovae \citep{Lo20}, thereby seeding the Universe with elements heavier than lithium \citep{LF13,Jo19}. This brings up the question: what is the minimum metallicity for planet formation? While the formation of giant planets is favored at higher metallicity \citep{WF15}, a similar trend is not apparent for terrestrial planets \citep{BBL14}. Theoretical models suggest that planet formation at metallicities $\lesssim 10^{-4}\,Z_\odot$ is not feasible \citep{JL13}, but this question is far from resolved. \\

\noindent \textbf{Bioessential elements:} Although carbon, hydrogen, nitrogen, oxygen, phosphorus, and sulfur (CHNOPS) are the canonical bioessential elements, life on Earth utilizes a very wide palette of elements in actuality, especially metals \citep{Mar16}. The precise abundances of bioessential elements in the high-redshift Universe is weakly constrained, but it is predicted that elements up to rubidium were synthesized by supernovae entailing high-mass stars \citep[Figure 1]{Jo19}. Thus, the availability of CHNOPS and certain other crucial elements (e.g., iron) might not be an issue. Models suggest that carbon-enhanced metal-poor (CEMP) stars and planets were present in the early Universe \citep{ML16}, and that the formation of water could occur at metallicities of $\sim 10^{-3}\,Z_\odot$ \citep{BSL15}. There are, however, some drawbacks that must be highlighted. Radioactive elements such as uranium would be rare due to the scarcity of neutron star mergers (which are responsible for their synthesis), thereby diminishing the prospects for radiogenic heating \citep[Figure 1]{Jo19}.\footnote{This point does not, however, rule out geothermal gradients since the latter are also driven by the primordial heat stemming from planet formation, which is a major contributor to the energy budget \citep{TS02}.} Molybdenum, which plays vital roles in biological functions \citep{Hi02}, is formed from dying low-mass stars or neutron stars \citep[Figure 1]{Jo19}, implying that it may exist in very low concentrations. \\

\noindent \textbf{Energy sources and gradients:} It is a well-known fact that living systems require the access to environmental gradients \citep{Kl16}. We will comment on electromagnetic radiation (photosynthesis) separately, but it bears noting that many other energy sources and gradients have been identified. For instance, both subaerial and submarine hydrothermal systems are characterized by steep gradients in pH, redox chemistry, and temperature \citep{RM04,SAB13,LL21}. More exotic founts of energy could also sustain life, such as magnetic fields, radioactivity, and tidal forces \citep[Chapter 5]{SMI18}. And last, but not least, the CMB would have contributed significantly to the energy budget at high redshifts. Given the diversity of energy sources, at least some of them may have been accessible on/in objects in the early Universe.\\

\noindent \textbf{Timescale for abiogenesis:} It must be noted at the outset that we do not have enough data to ascertain the timescale for the origin of life on other worlds \citep{ST12}. Even on Earth, we are confronted with a paucity for data. The earliest definitive biosignatures suggest that life was widespread on Earth a few hundred Myr after the inception of habitable conditions \citep{PTP18}. Phylogenetic methods suggest that life originated very shortly after the Moon-forming giant impact \citep{BPC18}, perhaps $\mathcal{O}(10)$ Myr after this event. Based on geological and chemical factors, \citet{LM94} argued that $\sim 7$ Myr could have sufficed for the evolution of oxygenic photoautotrophs \citep[cf.][]{Org98}. Taken collectively, it is not inconceivable that abiogenesis could occur on timescales of $\tau_L = 10$ Myr. Hence, if $\Delta t \gtrsim \tau_L$ is valid, the habitable epoch (embodied in $\Delta t$) might persist long enough for abiogenesis; from Table \ref{TableHabEp}, we notice that this criterion is fulfilled for most solvents. It is worth mentioning at this juncture that some worlds may possibly even witness the advent of technological intelligence in as little as $100$ Myr \citep{McK96,LL21}. \\

\noindent \textbf{Photosynthesis:} Life on Earth is deeply reliant on the access to low-entropy photons and the dissipation of energy as high-entropy photons \citep{Kl16}.\footnote{It has been theorized that the CMB could serve as the low-entropy source for planets orbiting black holes, which function as the high-entropy sinks \citep{ORB17}.} The most famous example is photosynthesis, whereby low-entropy photons from the Sun are converted into chemical energy and tapped for building biomass. The CMB is not well-suited for photosynthesis since most of the radiation is at mid-infrared (mid-IR) wavelengths and would therefore necessitate the evolution of sophisticated multi-photon schemes based on several photosystems coupled in series \citep{LL19}. If an object is, however, orbiting a host star, the latter will supply the low-entropy photons necessary for photosynthesis. 

In the habitable epoch of a given solvent, the chief advantage is that \emph{any} object around the host star -- as long as it is not situated too close -- permits the existence of that solvent and potentially habitable conditions for exotic life. This is beneficial because the minimum photon flux required for photoautotrophs is merely $\sim 10^{-5}$ of the incident photosynthetically active radiation (PAR) on Earth \citep{RKB00}. Hence, the maximum orbital radius ($a_\mathrm{max}$) at which photosynthesis is theoretically feasible is roughly two orders of magnitude higher than the outer limit of the conventional HZ \citep{LL19}. In fact, if life were to evolve means of concentrating PAR, then $a_\mathrm{max}$ may further increase by orders of magnitude \citep{Dy03}.

The question of whether photoautotrophy is capable of evolving over $\Delta t$ is obviously pertinent, and ties in with the preceding point. We note that genomic evidence favors an ancient origin for photosynthesis, indicating that it was prevalent by $3$ Ga or earlier \citep{SC20}. \\

\section{Discussion and Conclusion}\label{SecConc}
Motivated by the fact that the CMB energy density was much higher in the early Universe, we analyzed the epochs in which the CMB temperature was sufficiently high so as to permit the existence of appropriate solvents on the surface of any object. We calculated the redshifts (and ages) at which this epoch began and ended, as well as its duration ($\Delta t$), for different solvents. Our final results are presented in Table \ref{TableHabEp}.

Given, however, that this period of habitability occurs in a high-redshift Universe, there are many challenges that confront the origin and evolution of life. Hence, we provided a brief analysis of the salient factors. The issues that we tackled include planet formation, the access to bioessential elements, energy sources and gradients, and the time required for the origin of life. We suggested that some of these obstacles might not be as problematic as they seem at first glimpse, although there are several weakly constrained mechanisms at play.

On the basis of our analysis, one may ask which of the solvents in Table \ref{TableHabEp} is optimal inasmuch as harboring life in the early Universe is concerned. The solvent must satisfy the following list of desiderata.
\begin{enumerate}
    \item The epoch when the solvent can remain in liquid form ought to commence at $z \lesssim 70$ and ideally extend to as low redshifts as possible because the star formation rate and the number of terrestrial worlds ought to increase.
    \item A lower bound for $\Delta t$ might be $\sim 10$ Myr, when viewed from the perspective of abiogenesis, but it should preferably be comparable to $100$ Myr.
    \item The chemical species should be readily synthesized and available in sufficient cosmic abundance; for instance, multiple cases of detection in the interstellar medium would be indicative of the latter.
    \item There must be adequate empirical and/or theoretical grounds justifying why the solvent would be capable of facilitating biochemical reactions.
\end{enumerate}
As per Table \ref{TableHabEp} and the properties delineated in Sec. \ref{SecHW}, the three primary contenders are N$_2$, ethane, and supercritical H$_2$. The latter is manifestly the most well-suited insofar as criteria \#1 and \#2 are concerned, but there has been very little research with regards to its potential as a solvent (point \#4). While N$_2$ has the advantage of being liquid at lower redshifts compared to ethane and propane, this duo evince higher $\Delta t$ and there has been much more research undertaken in the realm of hydrocarbon solvents. Among these two options, propane is probably much less abundant than ethane in relative terms. Hence, of the various options investigated herein, ethane is probably the most promising contender.  

In summary, we have shown that the early Universe could have enabled the existence of different solvents and concomitant habitable conditions. Although the temporal interval would have been fairly brief, abiogenesis might have occurred on any object -- irrespective of its distance from a star -- by virtue of the temperature of the CMB. If life did originate in this period, it would have taken place at a time when the energy density of matter was much higher than that of dark energy (to wit, when $\Omega_m \gg \Omega_\Lambda$). This inequality would have major implications for the anthropic argument, which deals among other things with why we find ourselves in an era distinguished by $\Omega_m \sim \Omega_\Lambda$ \citep{Lo20}.

\acknowledgments
This research was supported in part by the Breakthrough Prize Foundation, Harvard University's Faculty of Arts and Sciences, and the Institute for Theory and Computation (ITC) at Harvard University.

%\bibliographystyle{aasjournal}
%\bibliography{EarlyUniv}

\end{document}